\PassOptionsToPackage{svgnames}{xcolor}
\documentclass[svgnames]{vgtc}                          

\usepackage{times}                     
\usepackage{amssymb}
\usepackage{tabu}                      
\usepackage{booktabs}                  
\usepackage{lipsum}                    
\usepackage{mwe}                       
\usepackage{amsmath}                    
\usepackage{mathptmx}                  
\usepackage{graphicx} 
\usepackage{epstopdf}
\usepackage{svg}
\usepackage{multirow}

\usepackage [english]{babel}
\usepackage [autostyle, english = american]{csquotes}
 \usepackage[final]{changes} 

\definechangesauthor[name={adder}, color=blue]{+} 
\definechangesauthor[name={deleter}, color=red]{-} 

\MakeOuterQuote{"}

\onlineid{1032}

\vgtccategory{Research}

\vgtcinsertpkg

\title{VR MRI Training for Adolescents: A Comparative Study of Gamified VR, Passive VR, 360$^{\circ}$ Video, and Traditional Educational Video}

\author{
Yue Yang\thanks{e-mail: yueyang1@stanford.edu, First Author.}\\ %
        Stanford University %
\and Mengyao Guo\\ %
     University of Macau %
\and Yuxuan Wu\\ %
     Stanford University %
\and Wally Niu\\ %
     Stanford University %
\and Emmanuel Corona  \\ %
     Stanford University %
\and Bruce Daniel \\ %
     Stanford University %
\and Christoph Leuze \thanks{e-mail: cleuze@stanford.edu}\\ %
     Stanford University %
\and Fred Baik \thanks{e-mail: fbaik@stanford.edu}\\ %
     Stanford University
     }

\abstract{ 
\textcolor{red}{Meta Quest Store: \url{https://www.meta.com/experiences/stanford-mri-simulator/8205539289482347/}}

Magnetic Resonance Imaging (MRI) can be a stressful experience for pediatric patients due to the loud acoustic environment, enclosed scanner bore, and a prolonged requirement to remain still. While sedation is commonly used to manage anxiety and motion, it carries clinical risks and logistical burdens. Traditional preparatory approaches—such as instructional videos and mock scans—often lack engagement for older children and adolescents. In this study, we present a comparative evaluation of four MRI preparation modalities: (1) a gamified virtual reality (VR) simulation that trains stillness through real-time feedback; (2) a passive VR experience replicating the MRI environment without interactivity; (3) a 360$^{\circ}$ first-person video of a real MRI procedure; and (4) a standard 2D educational video. Using a within-subjects design (N = 11, ages 10–16), we assess each method’s impact on head motion data, anxiety reduction, procedural preparedness, usability, cognitive workload, and subjective preference. Results show that the gamified VR condition has significantly lower head motion (p $<$ 0.001) and yielded the highest preparedness scores (p $<$ 0.05). Head motion data were significantly correlated with learning outcomes (p $<$ 0.01), suggesting that behavioral performance in VR strongly indicates procedural readiness. While all modalities reduced anxiety and were rated usable, interactive VR was preferred by most participants and demonstrated unique advantages in promoting engagement and behavioral rehearsal. We conclude with design recommendations for designing immersive simulations and integrating VR training into pediatric imaging workflows.
} 

\CCScatlist{
  \CCScatTwelve{mixed reality}{virtual reality}{magnetic resonance imaging}{MRI simulation};
  \CCScatTwelve{MRI simulator}{VR simulation}{adolescents anxiety}
}

\usepackage{cleveref}
\begin{document}

\teaser{
  \centering
  \includegraphics[width=\textwidth]{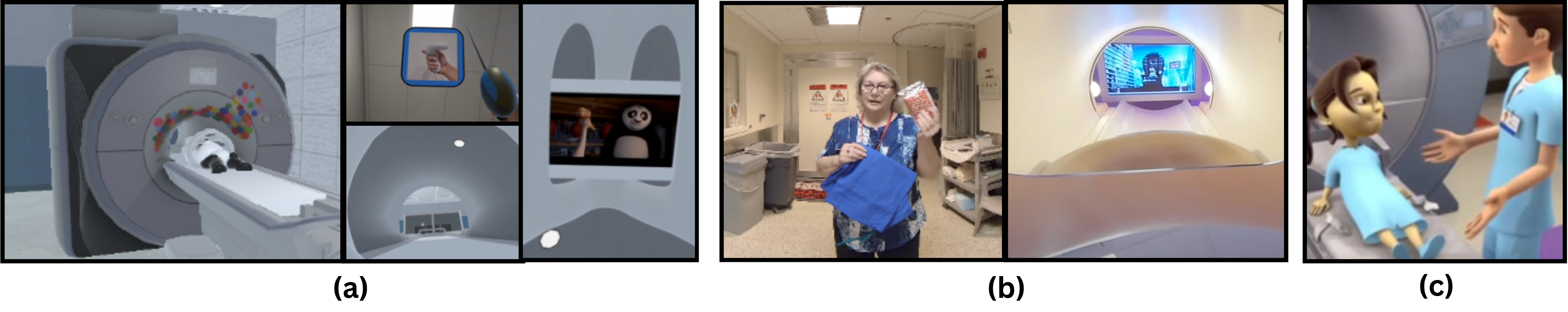}
  \caption{Overview of the four exposure methods used in our study. (a) A custom-designed VR MRI simulator featuring a panda avatar, a squeeze ball for interaction, and 6-DOF head tracking that aligns with a physical bed through passthrough capabilities. The simulator includes an optional interactive “hold still” game. This setup constitutes two distinct conditions—one with the game and one without. (b) A 360$^{\circ}$ video condition composed of two segments: the left panel presents an introductory scene captured by our 360$^{\circ}$ camera, while the right panel features a publicly available 360$^{\circ}$ video depicting the interior of an MRI scanner, including a movie display. These two segments together form the complete 360$^{\circ}$ session. (c) A widely circulated 2D cartoon video illustrating the MRI process, used as the baseline exposure condition.}
  \label{fig:teaser}
}


\firstsection{Introduction}

\maketitle


Currently, there are over 40 thousand MRI machines in the world and around 100 million MRI procedures are completed each year \cite{pmid30574259}. With increasing use, concerns about patient anxiety and claustrophobia have become more prominent, mainly due to the acoustic noise, claustrophobic environment, unfamiliarity of the procedure, the patient’s sense of control, and the requirement of holding still. Previous reports have shown that 1 to 15\% of all patients suffer from claustrophobia, resulting in over 2 million MRI procedures being terminated \cite{enders2011reduction}. During clinical screenings, up to 40\% of all patients report levels of anxiety, and acute anxiety prevention 2\% of scans from successful completion \cite{chapman2010mri, eshed2007claustrophobia, melendez1993anxiety}. If patients undergoing MRI feel particularly anxious or claustrophobic, they may actively ask the MRI technologist to stop the procedure or may passively require a rescan due to body movement and undiagnostic images. Among all MRI patients, pediatric patients are more sensitive to multiple MRI stress factors. It’s also particularly challenging for children to lie perfectly still for more than 30 minutes of an MRI scan and they cannot always see their parents in the scan room, causing feelings of discomfort, fear, and anxiety \cite{olloni2021pediatric}. More than 50\% of children also demonstrated negative behavior responses to medical procedures \cite{kain1996preoperative}. Specifically for MRI, previous reports have shown that the MRI termination rate of unsedated children ranges from 10\% (aged 6-17) to 47\% (aged 2-7 years) \cite{rosenberg1997magnetic, malisza2010reactions}. \par

To increase the success rate of MRI procedure, especially for pediatric patients, sedation via general anesthesia (GA) is generally required to achieve immobility. Moreover, the use of drug-induced GA has drastically increased over the last decade \cite{jung2020drug}.  One research group also reported that among >4500 pediatric patients undergoing MRI, 45\% of them require sedation \cite{stunden2021comparing}.  Although GA is highly effective in holding patients still, it has multiple limitations. Research has shown high levels of anesthesia-related mortality rates in developing nations (2.4-3.3 per 10,000 anesthetics), and children with developmental disabilities are 3 times more prone to hypoxia under sedation\cite{gonzalez2012anesthesia, kannikeswaran2009sedation}.  GA also increases anxiety levels for caregivers and parents due to the existence of potential negative reactions, prolonged recovery periods, and the application of drugs\cite{ayenew2020prevalence}. There is also a significant financial burden added to both patients and the healthcare system after applying GA to prepare patients for MRI. \par

Multiple MRI simulation methods have been proposed to avoid the limitations of GA, increase MRI success rate, and improve patients' overall feeling and experience undergoing MRI. Specifically, virtual reality (VR), a digitally developed 3D environment, was considered a cost-effective approach to effectively simulate the MRI experience for both adult and pediatric patients and generate comparable results with traditional hospital-based programs\cite{stunden2021comparing, nakarada2020can, jacob2023economic}.   Recent review papers also reported that VR has significant potential to prepare patients for MRI by reducing anxiety and claustrophobia, while suggesting the need for further software development to better replicate the real world and offer more engaging experience\cite{hudson2022scoping, cataldo2022use}. However, of the current studies, most of them used mobile headsets including MERGE VR and Google Cardboard that require a mobile device to run VR experience\cite{stunden2021comparing, ashmore2019free, brown2018virtual, liszio2020pengunaut}. Those mobile headsets are outdated and offer low-quality VR experience compared with more advanced and standalone VR devices. We also found only one study that used a more advanced headset \cite{garcia2007use}, the HTC VIVE, but still required a gaming computer to facilitate the VR experience and demanded significant time for setting up, which won’t be applicable to hospital settings. A recent paper compares VR MRI simulation with other traditional methods and 360$^{\circ}$ videos, but the study target adults and the mean user age is 30+ years old, making findings not generalizable to adolescents, who are reported to have higher termination and failure rates \cite{hosseini2025evaluating}. Thus, it is critical to understand how immersion help adolescents prepare for MRI.

\section{Related Work}
\subsection{Pediatric MRI Preparation \& Sedation}
Children and adolescents undergoing MRI procedures often experience heightened anxiety, which can lead to excessive motion, scan failure, or the need for sedation \cite{everts2022fear}. While sedation is almost universally required for infants and toddlers, older children and adolescents can often complete scans without pharmacological support if adequately prepared in advance \cite{thestrup2023nonpharmacological}. Given the risks associated with sedation—including respiratory depression, extended recovery, and the presence of an anesthesiologist—non-pharmacological interventions are widely preferred when feasible.

To mitigate anxiety and improve procedural compliance, many pediatric hospitals implement Child Life services, where trained specialists guide children through the imaging process using developmentally appropriate language, therapeutic play, and behavioral coaching \cite{olloni2021pediatric}. A common preparation tool within these programs is the pre-procedure instructional video, typically presented on a 2D screen. These videos explain the MRI process in simplified terms, describe the loud acoustic environment, and emphasize the importance of remaining still, all while reassuring children that the procedure is painless. Although these videos are cost-effective and broadly accessible, they remain passive in nature and may not adequately capture or sustain attention—particularly among adolescents who benefit from more interactive and immersive forms of engagement \cite{hogan2018video}.

\subsection{VR for Medical Procedure Preparation (Ages 10–18)}
VR has emerged as a promising modality for patient education and procedural anxiety reduction, particularly in pediatric care. By immersing users in simulated environments that approximate real-world scenarios, VR offers an experiential learning approach well-suited for exposure-based preparation and behavioral rehearsal. Prior studies have highlighted the efficacy of VR in reducing procedural distress among children, with applications spanning pain management, phobia treatment, and preoperative anxiety \cite{tas2022virtual, dehghan2019effect}.

In clinical contexts such as blood draws, vaccinations, and dental procedures, VR-based distraction—often delivered through interactive headset-based games—has been shown to significantly reduce self-reported pain and fear, outperforming conventional methods such as cartoons or passive video content \cite{sanchez2025effect}. These findings suggest that immersive, interactive content may more effectively capture attention and support pre-procedural preparation compared to traditional educational materials.

Within the domain of MRI preparation, several pilot studies and clinical prototypes have explored VR as a tool to familiarize pediatric patients with the scanning process \cite{ashmore2019free, van2023comparing}. These efforts aim to reduce anxiety and improve compliance by simulating the auditory and spatial experience of an MRI environment. However, variability in study design and limited use of control conditions make it difficult to generalize the observed benefits or directly compare outcomes with standard preparation methods. As such, further research is needed to systematically evaluate the comparative effectiveness of VR-based MRI preparation in adolescent populations.

\subsection{Gamification in Healthcare}
Gamification—the integration of game design elements such as goals, challenges, feedback, and rewards—has been widely recognized as a key mechanism for enhancing user engagement in digital health interventions. In pediatric populations, gamified experiences are particularly effective in sustaining attention, increasing motivation, and promoting active participation in therapeutic or educational activities \cite{cheng2019gamification}.

When applied to pre-procedural preparation, gamification can reframe a potentially intimidating medical scenario as a goal-oriented task, enabling children to feel more agency and control. For example, transforming the MRI preparation process into a game-like experience, where children play the game of playing-the-statue, can potentially reduce avoidance behaviors \cite{liu2024evaluating}. This interactive “learning-by-doing” approach aligns with developmental theories in pediatric psychology, which emphasize the importance of mastery experiences in fostering emotional resilience. Children cope better with medical stressors when they perceive success in completing a challenge, even within a simulated or imaginative context \cite{moore2006pretend}. By embedding these elements into immersive VR simulations, designers can create psychologically supportive environments that promote procedural understanding, reduce anticipatory anxiety, and enhance the child’s sense of preparedness.

 \subsection{Comparisons of 2D, 360$^{\circ}$ Video, and VR}
The increasing accessibility of consumer-grade VR and 360$^{\circ}$ video technologies has enabled new opportunities for immersive education and procedural training. Prior work has explored how different levels of media immersion impact user engagement, learning outcomes, and perceived presence \cite{im2023comparative}. Traditional 2D videos offer a familiar, low-barrier format for information delivery but lack spatial immersion and interactivity. In contrast, 360$^{\circ}$ video, particularly when viewed through a head-mounted display (HMD), allows users to explore the visual scene by looking in any direction, fostering a heightened sense of presence within the depicted environment. This modality has been utilized for applications such as virtual hospital tours, with studies showing that 360$^{\circ}$ video can help demystify clinical procedures \cite{ashmore2019free}. However, despite its immersive potential, 360$^{\circ}$ video remains a passive medium: users cannot alter the environment or interact with its content, limiting engagement and agency.

Comparative research across educational domains has shown that higher levels of immersion and interactivity, such as those afforded by full VR environments, can enhance engagement, improve retention, and strengthen the sense of presence \cite{makransky2021cognitive}. However, there is limited research on pediatric medical procedure preparation. Our work extends this line of research by directly comparing 2D video, 360$^{\circ}$ video, passive VR, and gamified \& interactive VR in the context of MRI preparation for older children and adolescents. By isolating media format as a variable and examining the role of intractability and gamification, we aim to provide empirical insights into the trade-offs between ease of access, psychological impact, and educational effectiveness across these formats.

\section{Method}
To investigate how varying levels of immersion and interactivity influence MRI preparation effectiveness in adolescents, we conducted a controlled, within-subject study comparing four preparation modalities:

\begin{enumerate} \item \textbf{Interactive VR Simulation}: A gamified VR experience featuring a virtual MRI scanner and an embedded “hold-still” game that trains users to minimize head movement during scan-like sequences. \item \textbf{Passive VR Experience}: The same virtual MRI simulation presented in a passive, non-interactive format, replicating the audiovisual environment without gameplay elements. \item \textbf{360$^{\circ}$ Video}: A first-person 360$^{\circ}$ video walkthrough of the MRI procedure, partially filmed at our institution and viewed through a HMD. \item \textbf{2D Informational Video}: A standard educational cartoon video about MRI procedures, presented on a flat screen using a conventional format. \end{enumerate}

\subsection{Participants and Design}
Eleven adolescents (N = 11; ages 10–16, $Median = 14$; 6 female, 5 male) participated in the study. All had normal or corrected-to-normal vision, no history of MRI experience, and no contraindications for VR use. Participants were screened for severe anxiety, claustrophobia, or developmental conditions that might interfere with the study. Informed assent and parental consent were obtained per IRB. To simulate real-world relevance, participants were told to imagine they “would soon need an MRI scan” and that each training module was intended to help them prepare.

We used a within-subjects design where all participants experienced four conditions: (1) interactive VR with a hold-still game, (2) passive VR, (3) 360$^{\circ}$ video, and (4) 2D video. Condition order was counterbalanced using a Latin square. Each session lasted 3–4 minutes, with short breaks and interim questionnaires between conditions. This design enabled direct comparisons within individuals while controlling for order effects. A post-study interview gathered reflections on overall preferences.

\subsection{Interventions and Preparation Modalities}
Each participant went through the following four preparation experiences:
\subsubsection{VR-MRI with Interactive Game (Game-VR)}
\begin{figure}[tbh]
\begin{centering}
    \includegraphics[width = \linewidth]{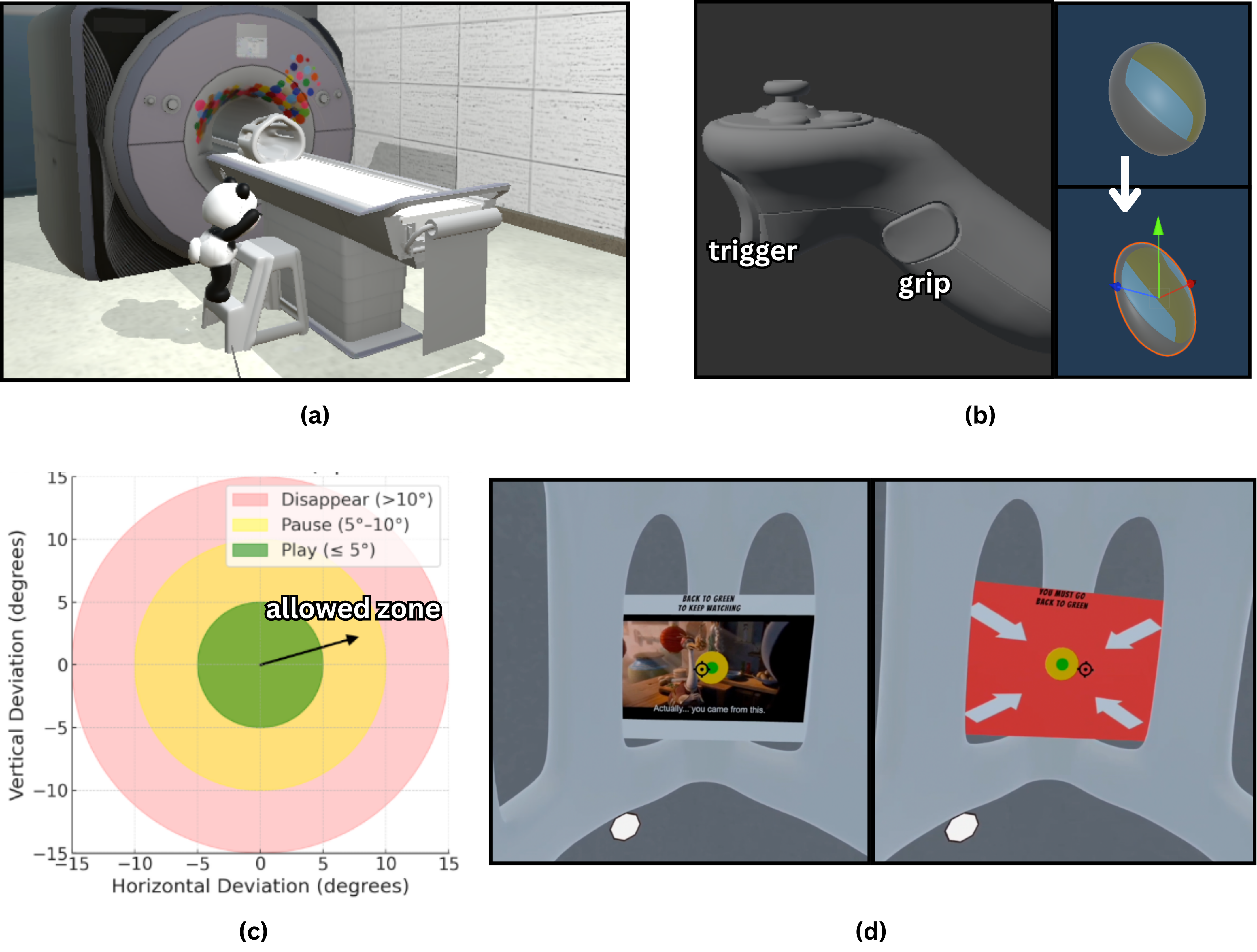}
    \caption{Gamified interactive VR experience. (a) The cartoon panda demo. (b) The squeeze ball feature. (c) Error threshold defined by our hold-still game. (d) Visualized HMD head pose that pauses or disappears the movie, corresponding with part c.}
    \label{fig:vrgame}
    \end{centering}
\end{figure}
Participants wore a VR headset (Quest 3) and were immersed in a simulated MRI environment that included a life-sized scanner, realistic hospital room details, and authentic MRI audio recorded from a clinical setting. The simulation incorporated a gamified element designed to reinforce the importance of staying still. As shown in \cref{fig:vrgame}, a cartoon panda avatar first introduced the task, explaining that participants must play a “statue game” and remain motionless when the MRI noise begins. Participants sit in a chair for this part of the game. Our simulation featured a head coil scan scenario, which is reported to elicit high levels of anxiety in pediatric patients.

Once the passthrough view starts, participants lie in a supine position on a physical bed to mimic the posture required during an actual MRI scan. In the virtual environment, they were then translated into the MRI bore. When the interactive hold-still game and MRI acoustic noise were launched, the system recorded the participant’s initial head orientation as a reference direction vector, denoted $\vec{s}$. Throughout the session, the participant’s instantaneous head direction was represented as vector $\vec{r}$. The angular deviation $\theta$ between the current orientation and the reference was calculated using the inverse cosine of the normalized dot product:

\[
\theta = \cos^{-1}\left( \frac{\vec{r} \cdot \vec{s}}{\|\vec{r}\| \cdot \|\vec{s}\|} \right)
\]

This angle $\theta$ determined the visual feedback through a video displayed on a virtual screen located 150~mm in front of the user in VR:

\begin{itemize}
    \item \textbf{Green zone:} $\theta \leq 5^\circ$ — The video \textit{plays continuously}.
    \item \textbf{Yellow zone:} $5^\circ < \theta \leq 10^\circ$ — The video \textit{pauses}.
    \item \textbf{Red zone:} $\theta > 10^\circ$ — The video \textit{disappears}.
\end{itemize}

To promote stillness, the participant’s head direction (head-pose ray) was continuously projected onto the screen, with a circular indicator at the center providing real-time feedback. When the head orientation remained within $5^\circ$ of the initial reference vector ($\theta \leq 5^\circ$), the indicator appeared green and the video played. Deviations between $5^\circ$ and $10^\circ$ triggered a yellow indicator and paused the video, while deviations exceeding $10^\circ$ resulted in a red indicator and caused the video to disappear. This feedback loop functioned both as a behavioral prompt and motivational mechanism, reinforcing sustained stillness through clear, immediate consequences.

To further simulate clinical realism and provide an opt-out mechanism, a virtual squeeze ball was rendered at the location of the participant’s right-hand controller. When both the grip and trigger buttons were pressed simultaneously, the squeeze ball visually shrank, and the participant was virtually withdrawn from the MRI bore. This interaction simulated early scan termination, mimicking how real patients can signal discomfort or request to stop the scan.

 \subsubsection{VR-MRI without Game (Passive-VR)}
 This condition also used the VR headset and the same virtual MRI room environment, but without any game mechanics. Participants lay down while wearing the headset and experienced a scripted virtual MRI scan simulation. They heard the MRI sounds and experienced a virtual scenario of an MRI scan, but no explicit visual feedback was given on their head motion. Essentially, this was a passive immersive experience with the goal of familiarizing the participant with the experience and sounds in an immersive way. Virtual environments for Game-VR and Passive-VR conditions were identical, allowing us to isolate the effect of the interactive gamification. Head motion data were also recorded during this session for analysis, though participants were not actively “penalized” for moving in this version.

\subsubsection{360$^{\circ}$ Video}
Participants viewed a pre-recorded 360$^{\circ}$ video depicting an MRI procedure from a first-person patient perspective. The 4-minute spherical video featured a typical MRI appointment, beginning in a radiology waiting room and progressing through interactions with a technician, lying on the MRI table, and experiencing the scan with realistic sound recordings. We recorded the technician portion at our institution and utilized a publicly available video for the experience portion \cite{360}, as in \cref{fig:teaser} (a) and (b). Participants wore Quest 3 to view the 360$^{\circ}$ content, allowing them to freely rotate their heads and explore the environment. Quantitatively, head motion was recorded during the scanning section. 

\subsubsection{Standard Training Video (2D Video)}
This was a conventional MRI educational video presented on a 2D screen. We used a publically available cartoon and reduced its duration to 2.5 minutes \cite{2D}. The video featured friendly animations and real MRI footage explaining how MRI works, what the child should do, and what to expect during the scan. It was presented on a 24-inch laptop screen placed on a table. The 2D video had play/pause control operated by the researcher, but participants mostly just watched passively. They were encouraged to pay attention and remember the key points, and no head movement data were recorded.

Each of the four sessions was roughly 3–4 minutes long. Including consent, instructions, breaks, and questionnaires, the total session per participant lasted about 1 hour. We hypothesize the following:
\begin{itemize}
    \item \textit{Reduced Head Motion with Game-VR.} Participants will exhibit the least head movement in the VR MRI simulation with the interactive hold-still game, since the game mechanics reward staying still. In contrast, we expect more movement in the other methods where there is no such gamified reinforcement.
    \item \textit{MRI Preparedness Correlates with Stillness.} We anticipate that participants who move less during the VR MRI sessions will score higher on preparedness measures. They will better remember and internalize the importance of holding still and other MRI procedures. In other words, effective engagement (as evidenced by staying still in VR) will be negatively correlated with head motion, indicating greater readiness for a real MRI.
    \item \textit{Anxiety Reduction.} We hypothesize that exposure to these preparatory experiences will reduce pre-study anxiety about MRI. In particular, Game-VR and Passive-VR conditions are expected to produce the largest drop in self-reported anxiety pre- and post-intervention.
    \item \textit {Overall Preference.} We predict that the Game-VR condition will be the overall preferred training modality. However, the usability score may be lowered due to the integration of additional interactive games. 
\end{itemize}

\subsection{Measures}
We collected a range of objective and subjective measures to evaluate each modality’s effectiveness
\subsubsection{Head Motion}
Head motion served as a key objective measure, functioning as a behavioral proxy for procedural cooperation and the ability to remain still during a simulated MRI scan. For the three immersive conditions—Game-VR, Passive-VR, and 360$^{\circ}$ Video—head position and orientation data were continuously recorded via the VR headset’s onboard sensors through Air Link using the Oculus Monitor program \cite{GITHUB}.

To quantify motion, we computed a composite motion index for each participant. One primary component of this index was the total time (in seconds) during which the participant's head deviated more than \(5^\circ\) from the baseline orientation vector. At each timestamp \(t\), we computed the angular deviation \(\theta_t\) using the dot product between the current head direction and the initial reference vector. The total time outside the stillness threshold, rounded to the nearest second, was calculated as:

\[
T_{\text{outside}} = \left\lfloor \sum_{t=1}^{T} \mathbb{1}[\theta_t > 5^\circ] \cdot \Delta t \right\rceil
\]

where \(\Delta t\) represents the sampling interval (\(1/60\) seconds). A lower \(T_{\text{outside}}\) value indicates better stillness and, by extension, improved procedural readiness.

\subsubsection{Anxiety}
We assessed participants’ anxiety levels before and after the first intervention using a questionnaire tailored for teens facing medical imaging. Before starting any training, participants completed a pre-study anxiety survey in which they rated how nervous they felt about having an MRI, including questions on feeling scared of the MRI, worried about lying still. We designed questions based on State-Trait Anxiety Inventory for Children (STAIC) and the Children’s Fear Survey Schedule. Each item was rated on a Likert scale (e.g., 1 = Not nervous at all, 5 = Very nervous). We aggregated these into a pre-intervention anxiety score. After only the first modality, participants answered a post-modality anxiety survey (retrospective post-test) asking them how anxious they now felt about an upcoming MRI, and how much the training helped reduce their fears. Reverse scoring was used for reversed scales. The difference between pre- and post-scores was used to evaluate \textit{Hypothesis 3}.

\subsubsection{Preparedness}
To evaluate how well each method taught the participant about the MRI procedure, we developed a Preparedness Questionnaire delivered after each session. This questionnaire included both objective and subjective items, and each question was used only once to avoid bias from participants anticipating repeated items.
1)  \textbf{Knowledge Problems}: e.g., “How long do you have to lie still during an MRI scan?” (open-ended), “Why is it important to hold still during MRI?” The questions covered different facts that were conveyed in all training modalities. We scored the number of correct answers as an indicator of knowledge retention: 2) \textbf{Reported Preparedness}: e.g., “On a scale of 1–5, how prepared do you feel to have an MRI after these training sessions?” and “Do you feel you could lie still in a real MRI scanner now?” These items measure the participant’s confidence and readiness. 3) \textbf{Sensory Preparedness}: specific questions, including “Do you remember what the MRI noise was like?” and “Can you describe how the MRI room looks?” to see if the modalities conveyed those sensory details effectively. For analysis, we report individual scores with the average result (preparedness) for each modality. \textit{Hypothesis 2} predicted a negative correlation between head motion and preparedness – the rationale being that those who were very still likely took the training seriously and learned more, or conversely, those who were unprepared tended to move more.

\subsubsection{System Usability and Cognitive Load}
After experiencing each modality, participants completed the Usability Metric for User Experience (UMUX-LITE) questionnaire. This fast, two-question questionnaire yields a score from 0 to 100 representing the overall usability and ease of use of a system. UMUX scores (above 80) are generally considered excellent usability. We hypothesized that all methods would be reasonably usable for this age group.

We further measured the perceived workload of each modality using the NASA Task Load Index questionnaire. Participants rated the mental demand, physical demand, temporal demand, performance (how successful they felt), effort, and frustration level for each training session. We included NASA-TLX because an overly demanding or stressful training could be counterproductive – e.g., if the VR game was too mentally taxing, it might stress the participant. We expected the standard video to result in a low workload, and the VR game to have higher mental or physical demand. By comparing TLX scores, we could evaluate if any method was notably harder or easier from the user’s perspective.

\subsubsection{Preference Ranking}
After experiencing all four methods, participants were asked to rank the four modalities from most preferred to least preferred as a preparation tool to address \textit{Hypothesis 4}. 

All quantitative measures (motion, quiz scores, Likert scales) were analyzed using appropriate statistical tests (details in Results). We also took observational notes during sessions (e.g., noting if a participant removed the VR headset early or showed signs of discomfort) and debriefed participants with an interview at the end, asking them to elaborate on their preferences and any suggestions.

\begin{figure}[tbh]
\begin{centering}
    \includegraphics[width =\linewidth]{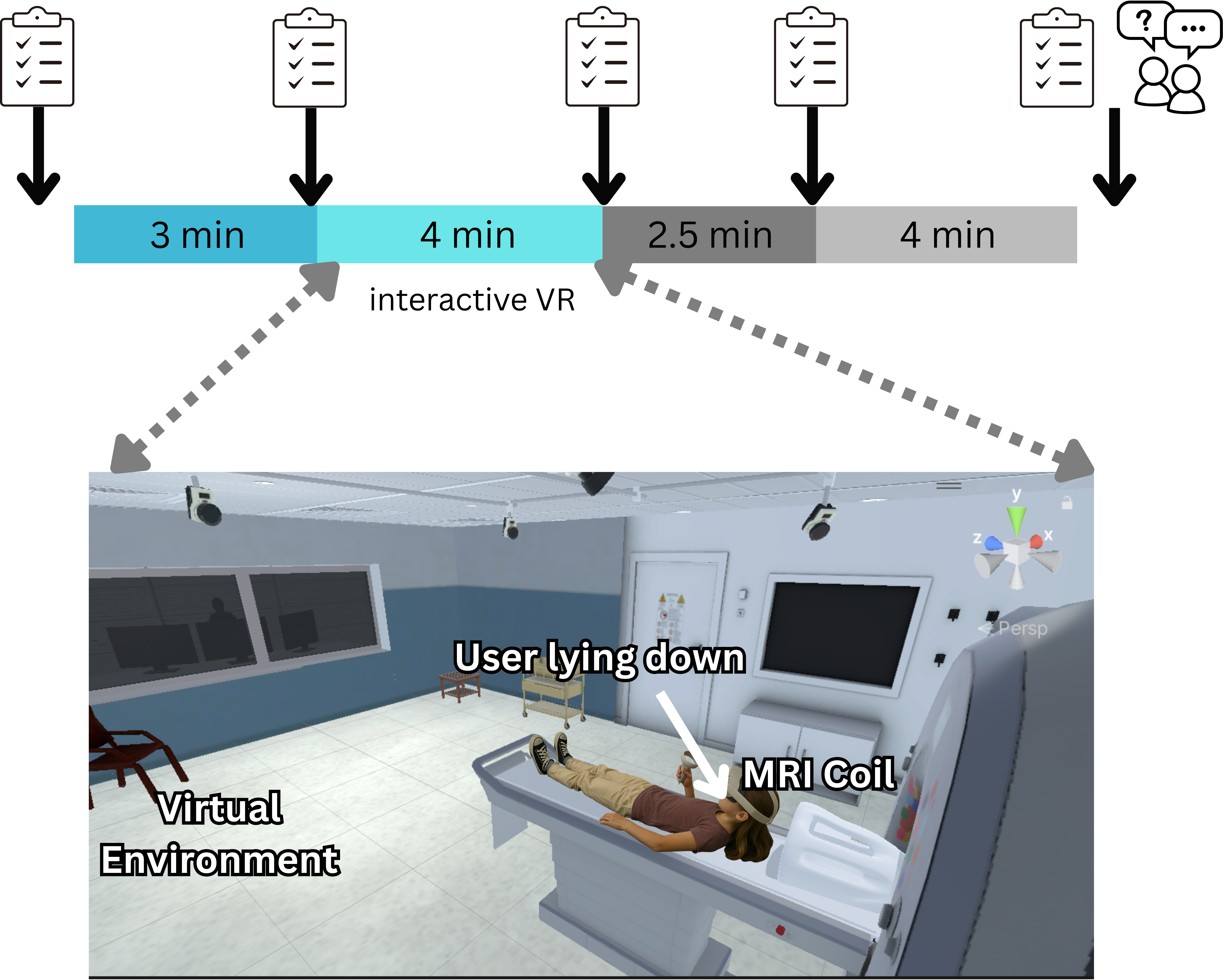}
    \caption{Structured study protocol. The figure illustrates the interactive Game-VR condition (from left to right: Passive-VR, Game-VR, 2D Video, 360$^{\circ}$ video). For illustration purposes, an image of a real participant has been overlaid onto the virtual environment. In the actual setup, participants lay on a physical bed aligned with the virtual MRI table shown in the VR simulation. }
    \label{fig:procedure}
    \end{centering}
\end{figure}

\subsection{Procedure}
As in \cref{fig:procedure}, each participant followed a structured protocol. Upon arrival, participants were introduced to the study and completed a brief demographics form. They were told they would try four different “training tools” designed to help prepare for an MRI. Before experiencing any modality, they completed a baseline anxiety questionnaire.

Participants then completed all four preparation modalities—interactive VR (Game-VR), Passive-VR, 360$^{\circ}$ video, and 2D video—in counterbalanced order. A researcher guided each session, assisted with headset placement. For VR sessions, participants lay supine while immersed in the simulation. For 2D and 360$^{\circ}$ video conditions, participants watched passively. Each condition lasted 3–4 minutes and was followed by a post-modality questionnaire. Participants took brief breaks between conditions. A brief semi-structured interview captured qualitative impressions after all four conditions. Motion data from the VR headset and headband sensor were timestamped and processed post-session to compute head motion indices. Questionnaire data were scored using standard methods. Analysis focused on insights using non-parametric statistics and effect sizes.

\section{Results}
In this section, we present the results of our study, organized by the main outcome measures: head motion, preparedness (knowledge and memory), anxiety, cognitive load and usability, and user preferences. Following non-normality test results, we used non-parametric tests (Friedman tests for overall differences and Wilcoxon signed-rank for pairwise comparisons). Significance was evaluated at $\alpha$ = 0.05 with appropriate corrections for multiple comparisons. We also report qualitative observations from participant feedback to enrich the quantitative findings.

\begin{figure}[tbh]
\begin{centering}
    \includegraphics[width = \linewidth]{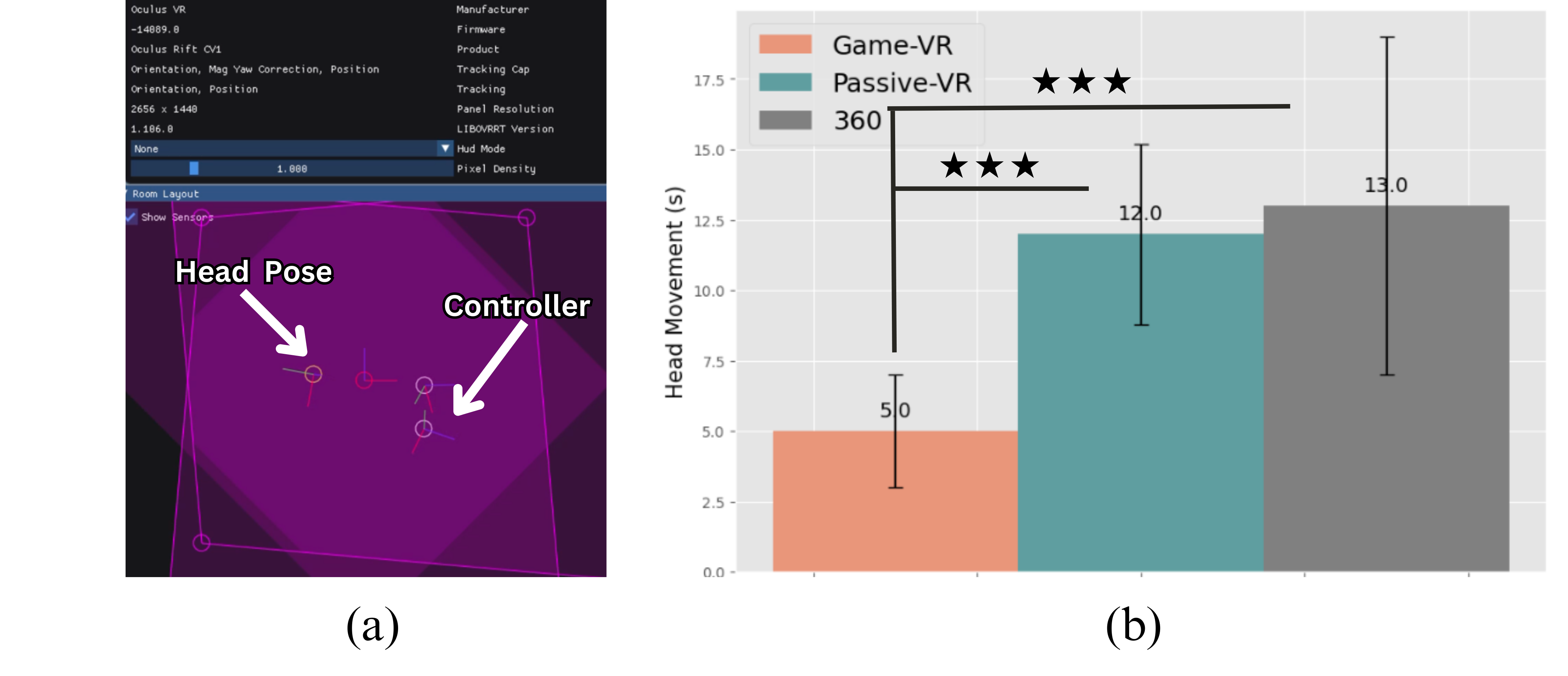}
    \caption{Results of head movement data recording and analysis. (a) The real-time tracking and recording of the camera and controller pose of Quest 3. The head pose was used for head movement analysis, and the controller pose was used for squeeze-ball mapping and rendering. (b) Median of head movement results (with standard error bars) observed in each experiment condition. Lower values indicate less movement (better stillness).}
    \label{fig:headmovement}
    \end{centering}
\end{figure}

\textbf{\textit{Head Motion Across Conditions.}}
The recorded head movement data supported our first hypothesis. Participants moved significantly less during the Game-VR condition than in any other condition. \cref{fig:headmovement} illustrates the median head motion for each preparation method, rounded to the nearest integer. A Friedman test found a significant effect of condition on head motion ($X^2$ = 26.0, p $<$ 0.001). Post-hoc comparisons confirmed that the Game-VR condition yielded lower motion than Passive-VR (p $<$ 0.01) and 360$^{\circ}$ Video (p $<$ 0.01). In contrast, the differences among the Passive-VR and 360$^{\circ}$ video were smaller and not statistically significant in pairwise tests (Passive-VR vs. 360$^{\circ}$: p $=$ 0.15). Meanwhile, in the passive-VR simulation, participants tended to move their heads more, potentially due to the lack of interactive hold-still game. The 360$^{\circ}$ video condition showed the highest motion on average, likely because participants naturally turned their heads to explore the full scene (some even sat up slightly at moments to get a better look around, as noted by the researcher).

This finding supports Hypothesis 1: the interactive gamification in VR clearly reduced motion. Many participants mentioned they were consciously trying not to move during the VR game, treating it like a challenge. In contrast, when simply watching the 360$^{\circ}$ video or the standard video, they “didn’t realize how much they moved.” One participant (P5) said, “In the game one, I was super focused on not moving at all. In the others, I was kinda looking around or didn’t think it mattered if I moved a bit.” Interestingly, some participants actually moved less under the 360$^{\circ}$ video condition than Passive-VR. One potential reason is that the 360$^{\circ}$ video, while immersive, was a real-world recording with limited dynamic stimuli, encouraging participants to remain relatively passive observers. In contrast, the Passive-VR simulation presented a fully rendered 3D environment, which may have invited more exploratory head movement—especially among participants curious to look around or test the boundaries of the virtual space. Without explicit feedback or a performance-based task, participants in the Passive-VR condition may not have felt a strong incentive to minimize motion.

Moreover, the fidelity of the VR environment itself may have influenced behavior. The stylized visuals and spatial audio in the VR simulation, though realistic, may have inadvertently prompted participants to reorient themselves or adjust their head position to follow virtual cues (e.g., sounds or character instructions). In contrast, the fixed, non-interactive camera in the 360$^{\circ}$ video constrained the perceptual experience, leading to fewer voluntary head movements. Nonetheless, both Passive-VR and 360$^{\circ}$ video conditions resulted in significantly higher motion than the gamified VR condition, underscoring the behavioral influence of interactivity and feedback.

Thus, these patterns highlight the importance of feedback and intentional design in training for motion-restricted procedures. Without a behavioral incentive or feedback mechanism—as in the Game-VR condition—participants may default to natural exploratory tendencies, especially when immersed in novel or visually engaging environments. It’s worth noting that no participant could stay perfectly still in any condition (there’s always some natural movement potentially due to the weight of HMD). We also checked if there was an order effect (maybe those who did Game-VR last would be practiced in staying still from prior sessions). We did not find a systematic order influence, regardless of when Game-VR occurred.

\begin{figure*}[tbh]
\begin{centering}
    \includegraphics[width = \linewidth]{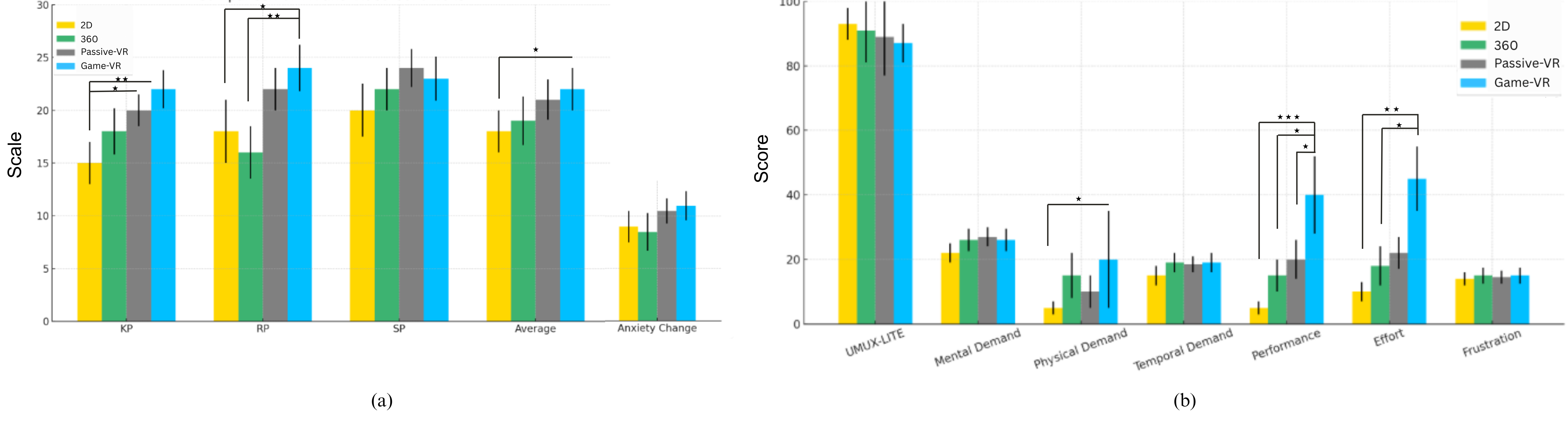}
    \caption{Results of user reported measurements. (a) Comparison of preparedness and anxiety for each modality. From left to right are Knowledge Problems (KP), Reported Preparedness (RP), Sensor Preparedness (SP), and the average of these three aspects to indicate the overall preparedness scale. (b) Anxiety scale change comparing baseline and post-modality results. Game-VR consistently shows high preparedness and anxiety scale compared to the other three conditions.}
    \label{fig:subjective}
    \end{centering}
\end{figure*}

\begin{figure}[tbh]
\begin{centering}
    \includegraphics[width = \linewidth]{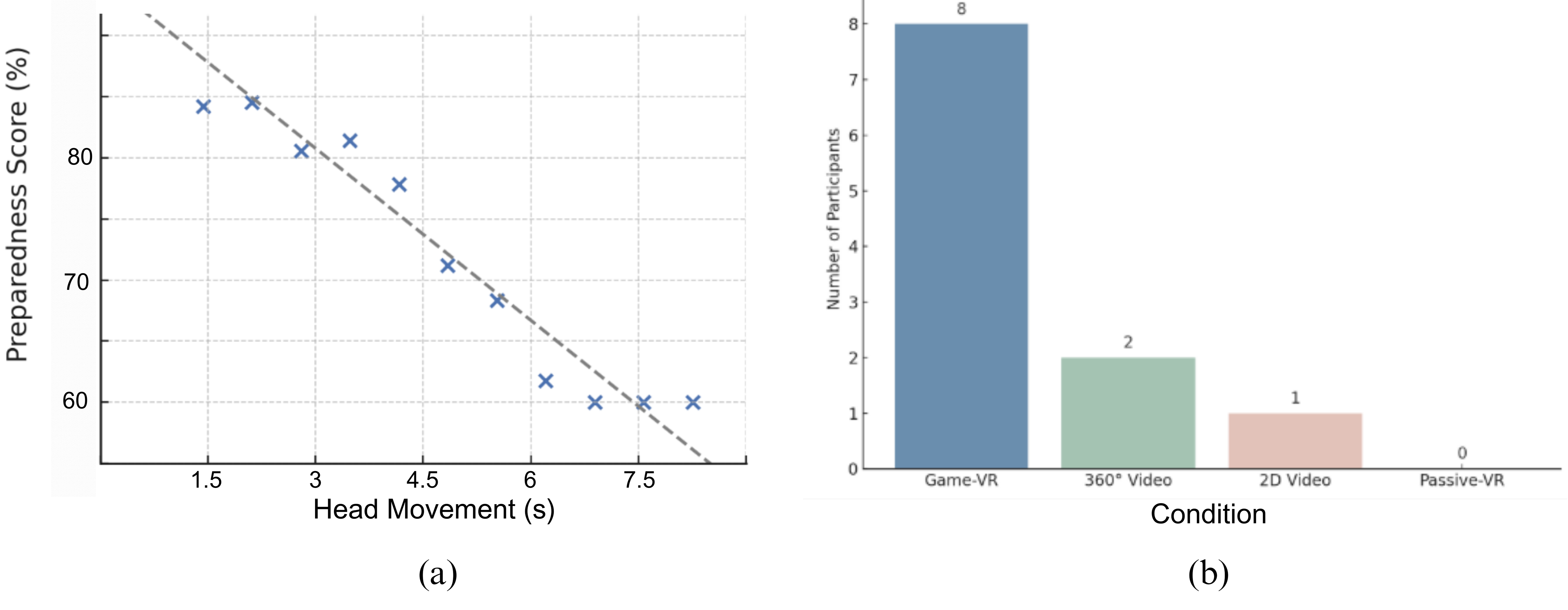}
    \caption{(a) Scatter plot of each participant’s preparedness score (out of 100\%) vs. head motion (s) in the Game-VR condition. A clear negative correlation is evident (dashed regression line), suggesting that participants who were more engaged in staying still (lower motion) also learned and internalized the material better, reporting higher preparedness for a real MRI. (b) Overall preferences reported by participants. Game-VR shows dominance over other conditions.}
    \label{fig:correlation}
    \end{centering}
\end{figure}

\textbf{\textit{Anxiety and Preparedness.}}
The average pre-study anxiety score (on a 5-point composite scale) was 3.5 (between “somewhat” and “moderately” anxious). As demonstrated in \cref{fig:subjective} (a), there is no significant difference between conditions in the anxiety reduction scale. The average post-intervention anxiety score, averaging all modalities, dropped to 2.6 (“a little” anxious). Thus, hypothesis 3 is only partially accurate. Exposure to preparation effectively decreased anxiety score, while immersive modalities show similar performance to 360$^{\circ}$ or traditional approaches. This suggests that collectively, the preparatory interventions helped alleviate fears about MRI.

Participants’ qualitative comments supported this reduction. Many reported that after going through the VR and videos, the MRI felt “more familiar” and “less scary” than they initially imagined. The loud noise was a major source of worry before; after hearing it in VR or video, most said it wasn’t as bad as they thought. For instance, P2 said, “When I first heard I might need an MRI, I was freaking out a bit. But now that I’ve seen what it’s like, I’m actually pretty calm about it.” We also looked at which modality participants credited the most for calming their nerves. This was often tied into their preference: those who loved the VR game often said it distracted them from anxiety by turning it into fun. P8: “The game one made me forget to be scared because I was playing and having fun watching movie.” Others found the immersive VR (with or without game) gave them a sense of experience that relieved the fear of the unknown. The passive-VR also got positive feedback for reassurance (“it felt like I already did an MRI, so I’m not as scared”). The 360$^{\circ}$ video was somewhat effective; some said seeing a real person go through it was comforting, but one participant found the 360$^{\circ}$ video “a bit dizzying,” which momentarily distracted from its reassuring value. The 2D video was informative, but a few said it was “boring” and didn’t really address their personal anxiety (“just info, not really calming, but useful” as P6 put it). In summary, all methods contributed to lowering anxiety.

For preparedness, both the Game-VR and Passive-VR conditions demonstrated notable advantages over the 2D video modality. Specifically, participants showed significantly higher performance on KP tasks—such as accurately describing the characteristic MRI scanner noise or explaining the rationale for remaining still—following exposure to the Game-VR (p $<$ 0.01) and Passive-VR (p $<$ 0.05) modalities. These findings suggest that immersive experiences are more effective at reinforcing factual retention, likely due to their ability to simulate procedural context, thereby promoting deeper cognitive encoding.

In contrast, when examining RP problems, significant differences were observed between the Game-VR condition and both the 360$^{\circ}$ video (p $<$ 0.01) and 2D video (p $<$ 0.05) conditions. Interestingly, while the 360$^{\circ}$ video was designed to provide a realistic first-person perspective, its median RP scores were lower than those of the 2D video, though the difference did not reach statistical significance. One plausible explanation is that some participants reported mild discomfort or dizziness while navigating the 360$^{\circ}$ video, which may have distracted from content absorption. Additionally, the passive nature of the 360$^{\circ}$ modality, despite its visual realism, may not have sufficiently encouraged active engagement or internalization of procedural strategies. With respect to sensory preparedness, defined as the ability to recall and describe the ambient environment and auditory cues of the MRI suite, no significant differences were found between modalities. This suggests that all conditions, including the 2D video, effectively conveyed the basic sensory expectations of the procedure, such as the scanner noise and room setup. Overall, these findings provide evidence that the Game-VR condition not only outperforms the 2D video in conveying factual information but also fosters greater preparedness across multiple dimensions. When averaging across KP, RP, and SP scores, the Game-VR condition resulted in significantly higher overall preparedness than the 2D video (p $<$ 0.05), revealing the value of interactivity and game design in procedural education.

Crucially, to further evaluate Hypothesis 2, we examined the correlation between participants’ head motion (in Game-VR) and their preparedness score (composite of quiz and self-rating). We found a strong negative correlation (Spearman’s $\rho \approx -0.75$, p $<$ 0.01), supporting Hypothesis 2 (see \cref{fig:correlation} (a)). In other words, those who managed to stay very still during the VR game tended to have higher preparedness scores (they answered more knowledge questions correctly and felt more confident about the MRI), whereas those who moved more tended to have lower preparedness outcomes.

It’s important to note that correlation does not prove causation: possibly those who were inherently more motivated or less anxious paid more attention both to not moving and to absorbing information. Nevertheless, the result aligns with our expectation that the act of practicing stillness (with feedback) reinforced the importance of the practiced skill. Several participants commented that the VR game taught them why staying still matters. For instance, P3 said, “The game made me realize every little movement counts. In a real MRI, that probably means blurry pictures. Now I know I really have to focus on not moving.” Specifically, when asked “Why do you need to hold still in an MRI?” all 11 participants answered correctly after the sessions.

\textbf{\textit{Usability and Cognitive Load.}} 
Demonstrated in \cref{fig:subjective} (b), all four preparation modalities were rated favorably on usability, as measured by the UMUX-LITE. Median UMUX scores (on a 0–100 scale) were as follows: 2D Video, 92; 360$^{\circ}$ Video, 90; Passive-VR, 88; and Game-VR, 85. A UMUX score above 72 is generally considered to indicate good usability, suggesting that all conditions met or exceeded usability expectations for adolescent participants. Notably, the 2D video was rated highest in usability, likely reflecting its simplicity and participants' familiarity with traditional screen-based content. Although Game-VR received the lowest median usability score among the modalities, it still fell within the “excellent” range. Several participants attributed their slightly lower Game-VR scores to initial discomfort or confusion—for instance, one commented that “the headset was a bit heavy,” while another noted that “it took a moment to figure out what to do in the game.” Usability scores for Passive-VR showed greater variability. One participant assigned a score of 70 due to uncertainty during the initial moments of the simulation, stating, “I wasn’t sure if I was doing it right at first.” However, once participants adjusted, most found the VR-based modalities intuitive to use. These findings suggest that while immersive technologies may introduce a short learning curve, adolescents are capable of adapting to them quickly with minimal guidance.

We further collected cognitive workload data using the NASA-TLX, which captures six dimensions of perceived workload: mental demand, physical demand, temporal demand, performance, effort, and frustration. We observed modality-specific differences in cognitive load that aligned with the nature of the interaction. The 2D video condition yielded the lowest overall workload scores, with participants reporting negligible mental or physical effort. Engagement with this modality was largely passive, and participants described the task as easy to follow. The 360$^{\circ}$ video condition presented a slightly higher workload (mean $\approx$ 18/100), largely due to the need for active head movement to explore the immersive video space and the minor physical effort required to stabilize the headset. Some participants noted this additional demand but did not find it burdensome.

Passive-VR exhibited a comparable workload level to 360$^{\circ}$ video (mean $\approx$ 20/100), with participants reporting low-to-moderate mental effort to remain still and understand the virtual procedure. The Game-VR condition incurred the highest cognitive load among all modalities (mean $\approx$ 30/100), as it required sustained attention, self-monitoring of motion, and effort to “win” the hold-still challenge to keep the movie playing. Several participants described the experience as “I was trying to stay not to move, so it was somewhat hard,” while a small subset noted mild frustration when inadvertently triggering in-game feedback penalties due to movement. Importantly, this frustration remained within acceptable limits—the highest frustration score for Game-VR was 27/100, attributed to initial confusion about gameplay mechanics. One participant also reported higher physical demand (60/100) due to discomfort from lying still with the headset on for an extended period.

A Friedman test revealed significant differences in overall workload across the four conditions (X² = 24.5, p $<$ 0.001). Post-hoc pairwise comparisons showed that participants reported significantly higher performance for Game-VR compared to 2D video (p $<$ 0.001), 360$^{\circ}$ video (p $<$ 0.05), and Passive-VR (p $<$ 0.05). Moreover, Game-VR was rated significantly higher in effort compared to 2D video (p $<$ 0.01) and 360$^{\circ}$ video (p $<$ 0.05). The 2D video also showed significantly lower physical demand than Game-VR (p $<$ 0.05). These findings confirm that the interactive nature of Game-VR inherently requires greater cognitive efforts and physical engagement. However, the workload levels, while higher than those of passive conditions, remained well within a tolerable and manageable range for all participants. This suggests that while Game-VR introduces a modest increase in effort, it does not compromise usability, indicating a balance between challenge and user comfort (an essential consideration in pediatric patient education design).

\textbf{\textit{User Preferences.}} 
Following the completion of all four preparatory modalities, participants were asked to indicate which method they found most helpful and which they would personally choose if preparing for a real MRI scan. As shown in \cref{fig:correlation} (b), a strong majority (8 out of 11) selected the interactive VR with hold-still game (Game-VR) as their preferred method. Two participants chose the 360$^{\circ}$ video as their top preference, while one participant favored the standard 2D educational video. Notably, no participant identified the Passive-VR simulation as their preferred option.

The participants who selected Game-VR consistently emphasized its interactivity and behavioral engagement as key advantages. Many described the experience as “fun,” or “helpful.” The 360$^{\circ}$ video was appreciated by two participants who valued its use of real-world imagery and the ability to explore the MRI setting from a first-person perspective. These participants mentioned that seeing the actual process made the experience feel more realistic. However, some participants reported minor discomfort such as dizziness or difficulty maintaining attention during the 360$^{\circ}$ session. 

Only one participant selected the 2D educational video as their top choice. This individual expressed a preference for traditional instructional formats and reported feeling more comfortable with the cartoon explanations and factual narration. While this modality lacked the immersion and interactivity of VR-based approaches, it was still recognized as useful by several participants, particularly in conveying procedural expectations. Across the cohort, participants acknowledged the value of combining multiple preparation methods. Several expressed that pairing the VR game with a brief 360$^{\circ}$ video could be fun. This suggests that a hybrid preparation approach may offer the most comprehensive benefit for diverse adolescent users. These findings offer further support for Hypothesis 4, confirming that the gamified VR experience was perceived as the most preferable preparation method among the modalities tested. The absence of preference for the Passive-VR condition highlights the importance of interactivity and gamification in fostering both engagement and learning, particularly among adolescent users.

\section{Discussion and Conclusion}

Our study evaluated the effectiveness of four distinct MRI preparation modalities for adolescents, with a particular focus on interactivity and immersion. The findings provide compelling evidence that Game-VR outperforms Passive-VR, 360$^{\circ}$ video, and traditional 2D educational video in promoting procedural stillness, enhancing preparedness, and fostering user preference. These results have important implications for the design of pediatric MRI preparation interventions, particularly in populations where anxiety and motion are barriers to successful imaging.

\subsection{Behavioral Engagement Through Gamified Interactivity}

The clearest advantage of the Game-VR modality was its significant reduction in head motion during the simulated scan. This supports the hypothesis that active behavioral engagement, reinforced through real-time feedback, can effectively train stillness—a critical skill for MRI compliance. Importantly, we found a strong negative correlation between head motion and preparedness scores, suggesting that behavioral focus during the simulation translated into cognitive learning and confidence. These findings align with prior literature on experiential learning and support the idea that immersive, goal-oriented tasks can drive both psychological and behavioral readiness in medical contexts.

\subsection{Subjective Preparedness and Anxiety Reduction}

All four modalities contributed to reduced self-reported anxiety following exposure, indicating that preparatory interventions—regardless of delivery format—can play a valuable role in alleviating anticipatory stress. However, interactive VR provided additional benefits in fostering perceived preparedness, particularly in helping participants understand the importance of stillness and feel equipped for the actual MRI environment. The immersive VR conditions (Game-VR and Passive-VR) facilitated better recall of procedural details and improved self-reported readiness compared to passive video-based approaches. This suggests that immersive modalities can promote deeper cognitive encoding and contextual understanding, likely due to the multi-sensory experience and spatial congruence with the real-world procedure.

\subsection{Trade-offs in Cognitive Load and Usability}

Although the Game-VR condition imposed the highest cognitive and physical demand among the four modalities, this increased workload did not compromise usability. All modalities scored within the “good” to “excellent” usability range on the UMUX-LITE, and most participants found the VR interface intuitive after minimal adjustment. Notably, the 2D video achieved the highest usability score, reflecting its simplicity and familiarity. However, its passive nature likely limited its effectiveness in preparing users for a behaviorally demanding task like MRI. These results indicate that while gamified VR introduces modest increases in effort, it remains accessible and acceptable for adolescent users when carefully designed.

\subsection{Implications for Pediatric MRI Preparation}

The majority of participants preferred the Game-VR modality, emphasizing its interactivity, feedback mechanisms, and game-like structure. This underscores the importance of designing interventions that go beyond passive education, particularly for older pediatric patients and adolescents who may benefit more from active participation. The lack of preference for Passive-VR further highlights that immersion alone is insufficient without task engagement. Interestingly, the 360$^{\circ}$ video—despite its realism—elicited mixed responses due to potential discomfort and lack of agency, suggesting that realism must be balanced with interactivity and user comfort.

From a deployment perspective, the use of standalone VR headsets like the Meta Quest 3 offers a practical advantage over prior VR implementations that relied on external PCs or mobile-based systems. The minimal setup time and portable nature of modern VR devices make them viable for integration into clinical workflows, potentially reducing reliance on sedation and improving patient outcomes.

\subsection{Limitations and Future Work}

Several limitations must be acknowledged. First, the sample size (N=11) limits the generalizability of our findings. While the within-subjects design enhances statistical power, future studies should replicate these findings in larger and more diverse cohorts, including children with previous MRI experience or diagnosed anxiety disorders. Second, while our composite motion metric provides a useful proxy for stillness, it may not capture all aspects of procedural readiness. Physiological measures such as heart rate variability, or long-term outcomes like actual scan success rates, could offer complementary insights. Third, while our study used a short exposure duration (~3–4 minutes), future work could explore the impact of longer or repeated sessions on learning retention and anxiety resilience.

Additionally, hybrid modalities that combine multiple features (e.g., an introductory 2D video followed by a VR simulation) may prove beneficial for diverse learners. Adaptive difficulty mechanisms and personalized feedback could further enhance training efficacy, particularly for children with varied cognitive or sensory profiles.

\subsection{Conclusion}
This study demonstrates that a gamified VR MRI simulator can significantly improve adolescents’ behavioral and cognitive readiness for MRI compared to passive or video-based approaches. By training stillness through interactive feedback, reducing anxiety, and promoting preparedness, such tools have the potential to reduce the need for sedation and improve procedural success rates. Our results support integrating immersive, interactive technologies into pediatric imaging workflows, and highlight the importance of thoughtful design in creating developmentally appropriate digital health interventions. As VR technology becomes increasingly accessible, its application in pediatric procedural preparation warrants continued exploration and clinical translation.

\subsection{Acknowledgments}
The authors wish to thank all study participants and MRI technologists Kevin Epperson and Karla Epperson who assisted in video recording. The authors appreciate feedback received from the Chariot program and the Lucas Center at Stanford Medicine.

\bibliographystyle{unsrt}

\bibliography{template}
\end{document}